# A Law of Nature?

Marvin Chester

chester@physics.ucla.edu

Physics Department, University of California, Los Angeles

Abstract:

Is there an overriding principle of nature, hitherto overlooked, that governs *all* population behavior? A single principle that drives all the regimes observed in nature: exponential-like growth, saturated growth, population decline, population extinction, oscillatory behavior? In current orthodox population theory, this diverse range of population behaviors is described by many different equations - each with its own specific justification. The signature of an overriding principle would be a differential equation which, in a single statement, embraces all the panoply of regimes. A candidate such governing equation is proposed. The principle from which the equation is derived is this: *The effect on the environment of a population's success is to alter that environment in a way that opposes the success.*

## 1. Introduction

What are the conceptual foundations in ecological studies? Are there laws of nature governing ecological systems? What are they?

Over the years there has grown a community of scholars who have grappled with these profound questions. (Murray 2000, Murray 2001, Turchin 2001, Ginzburg and Colyvan 2004, Lange 2005, Gorelick 2011, Colyvan and Ginzburg 2010)  Some have concluded that concern for such laws is not the business of ecological research. (Egler 1986, O'Hara 2005)  Others have concluded that Malthusian exponential growth constitutes an essential law (Berryman 2003).  That exponential growth and the logistics equation are of great conceptual utility but *not laws* of nature is argued cogently by still others.  (Lockwood 2008, Holt 2009)

Focussing on the lofty notion, Law of Nature, may be a distraction from a more elemental pursuit: to understand nature. That is surely the goal of science. Progress in understanding is marked by conceptual coalescence: the quest to embrace an ever larger body of findings with ever fewer statements of principle. Paraphrasing Mark Twain, the task of science is to describe a plethora of phenomena with a paucity of theory.

Newton showed that the motion of things on earth are governed by the same rules as the motions of heavenly bodies. Formerly these two had appeared to be unrelated domains. Newton showed that a single principle governed them both. This synthesis was magnificently fruitful. It underlies our understanding of anything mechanical or structural. Much of our material well being depends on it.

Darwin's principle of natural selection explained a wealth of biological phenomena by a single idea. Through his synthesis the concept of evolution became part of our intellectual heritage.

Wegener showed us that continental drift - plate tectonics - is the underlying reason for such diverse phenomena as the distribution of fossils in the world, the shape of continents, earthquake belts and volcanic activity. That insight has proved remarkably beneficial.

Mendeleev gave structure to the chaos of chemistry with his table of the elements. He consolidated a profusion of chemical data into an all encompassing tabular statement of principle. This undertaking led to the understanding that matter was made of atoms. (Mendeleev, himself, never believed this!)

James Maxwell brought electricity and magnetism together by an overriding formalism that covered them both. The undertaking gave rise to an understanding of the nature of light.

Laws of Nature are not immutable. They may lose their status. This process of conceptual coalescence is an ever evolving one. Newton's laws on mechanical motion and Maxwell's on electromagnetism are incompatible. In 1905 Einstein produced a theory that embraced both of these vaste domains. In it Newton's principles become a limit behavior of a more all inclusive theory - relativity. So a law of nature can be dethroned - albeit still cherished and useful. It can be subsumed under a principle which embraces a larger domain of phenomena. The broader the scope of applicability the more valuable is the theory. Einstein's laws of nature absorbed Newton's. Relativity spawned nuclear energy, a deeper understanding of stellar processes and much more.

All of these examples have in common that a wide breadth of empirical observation is accounted for by a single idea. We see in them that conceptual coalescence is a foundation stone of scientific understanding. In that spirit, offered here is a candidate synthesis: a single equation that brings together the disparate domains of population behavior. We suggest that the panoply of population behaviors all issue from a single principle.

In current orthodox population theory, the diverse range of population behaviors is described by many different equations - each with its own specific justification. Every regime has its own special theoretical rationale. Exponential growth has a limited range of validity. The Logistics Equation describes another restricted regime. Oscillatory behavior demands that a new paradigm be requisitioned; the Lotka-Volterra equations (Lotka 1956, Volterra 1926) or, because their solutions are not structurally stable, their later modifications (Murray 1989, Vainstein et al 2007). And none of these describe population decline, nor population extinction. Contemporary theory offers no overriding principle that governs the entire gamut of population behaviors.

As long ago as 1972 (Ginzburg 1972), in a challenge to orthodox convention, L. R. Ginzburg took the bold step of proposing that population dynamics is better represented by a second order differential equation. All accepted formulations relied on first order differential equations as they still do today. He developed his thesis over the years (Ginzburg 1986, 1992, Ginzburg and Taneyhill 1995, Ginzburg and Inchausti 1997) culminating in the pithy and persuasive book, "Ecological Orbits" (Ginzburg and Colyvan 2004).

Suppose the family of solutions to a single second order equation should match population behavior just as well as the many accepted first order equations do. That family of solutions have in common their single progenitor. Embedded in them is the principle that generated them.

When the family of solutions to a differential equation is found to fit empirical reality then that equation is expressing a truth about nature. It can give us insights and enable us to make predictions. Producing a second order equation whose solutions characterize a variety of population behavior is equivalent to uncovering a principle of nature governing those populations.

In the following we take a route different from Ginzburg's and arrive at a substantially different equation - albeit a second order one. We procede from a guess at what may be the underlying principle and then derive the second order differential equation that expresses that principle. If empirical reality is well fit by the progeny of that equation then we may conclude that the principle is true. And we will have produced a conceptual coalescence: a tool for better understanding nature.

## 2. Traditional Perspective

Call the number of members in the population, n. At each moment of time, t, there exist n individuals in the population. So we expect that n=n(t) is a continuous function of time.

The rate of growth of the population is dn/dt; the increase in the number of members per unit time. That this is proportional to population number, n, is the substance of Malthus' idea of 'increase by geometrical ratio'. Call the constant of proportionality, R. Then the well known differential equation that embodies the idea is:

$$dn/dt = Rn \qquad (1)$$

It is a first order differential equation and when R is constant, its solution yields the archetypical equation of exponential growth.

Now, common experience tells us that exponential growth cannot proceed indefinitely. "Most populations do not, in fact, show exponential growth, and even when they do it is for short periods of time in restricted spatial domains," writes R. D. Holt (Holt 2009). No population grows without end.

The first efforts to expand the breadth of applicability of theory to observation - to acheive some conceptual coalescence - was to allow R to vary with time.  The motivation was to retain that appealing exponential-like form and seek to explain events by variations in R. "The problem of explaining and predicting the dynamics of any particular population boils down to defining how R deviates from the expectation of uniform growth" (Berryman 2003). The concept is that exponential growth is always taking place but at a rate that varies with time. The idea is ubiquitous in textbooks. (Britton 2003, Murray 1989, Turchin 2003, Vandermeer 1981)

An object example of this process is provided by the celebrated Verhulst equation.

$$\frac{dn}{dt} = nR(n) = nr(1 - \frac{n}{K}) \qquad (2)$$

Here the constant, r, is the exponential growth factor and K is the limiting value that n can have - "the carrying capacity of the environment" (Vainstein *et al.* 2007). The equation insures that n never gets larger than $n_{MAX}$ = K. A population history, n vs. t, resulting from this first order differential equation is the black one of Figure 1.The Verhulst equation - often cited as the Logistics Equation - is regularly embedded in research studies. (Nowak 2006; Torres e*t al.* 2009; Jones 1976; Ruokolainen *et al.* 2009; Okada *et al.* 2005; Ma 2010)

## 3. Shortcomings of the traditional perspective.

The textbook mathematical structure outlined in the last section has acquired the weight of tradition. Keeping the exponential-like form by allowing R to vary is certainly appealing. But it has this serious failing: the practice forbids description of several known regimes of population behavior. It denies further conceptual coalescence. For example, unless R is taken as imaginary the observation of population oscillations cannot be described in this formalism.

Another proscribed regime is extinction. A phenomenon well known to exist in nature is the extinction of a species. "... over 99% of all species that ever existed are extinct" (Carroll 2006). But there exists no finite value of R - positive or negative - that yields extinction! It cannot be represented by R except for the value negative infinity; $-\infty$. So, in fact there is good reason to avoid R as the key parameter of population dynamics.

In the continuous-n perspective the mathematical conditions for extinction are these: n=0 and dn/dt<0. No infinities enter computations founded on these statements. Hence embracing n(t) itself as the key variable directly allows one to explore the dynamics of extinction.

Next consider the eponymous Verhulst Equation (the Logistics Equation). As Verhulst himself pointed out (Verhulst 1838) it is motivated only by the observation that populations never grow to infinity. They are bounded.

But there are other ways - not describable by Verhulst's equation - in which population may be bounded. For example, n(t) may exhibit periodicity. Or, as in Figure 1, a curve essentially the same as Verhulst's may arise from an entirely different theory where no K = $n_{max}$ limit exists. One of the possible population histories resulting from the alternative theory offered below – which contains no $n_{MAX}$ - is shown in blue. Data fit by one curve will be fit just as well by the other. The limited validity of r/K Selection Theory has been noted by researchers over the years. (Parry 1981; Kuno 1991)

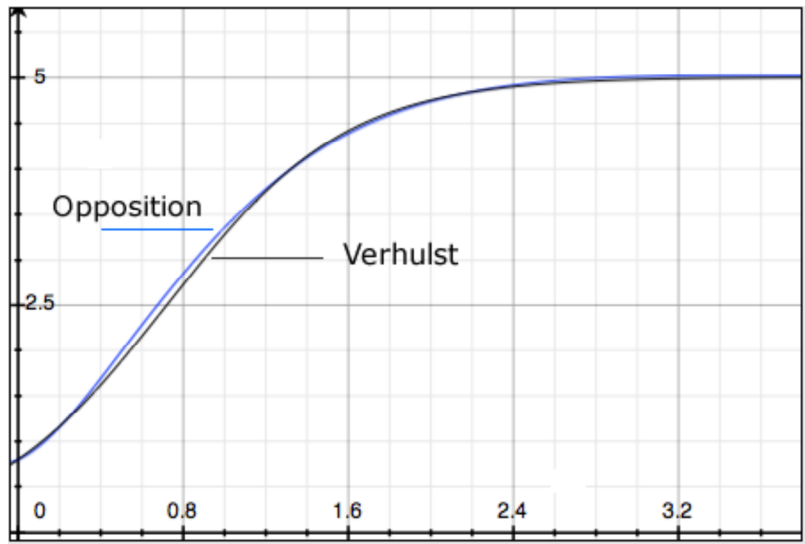

Figure 1. Two population histories: number vs. time. The black curve is the Verhulst (Logistics) Equation. The blue curve is one of the solutions to the Opposition Principle differential equation (5.6). Where one curve fits data so will the other.

Thus the accepted Malthusian Structure of population dynamics has, and will always have, only a limited domain of validity. Many population histories cannot be fit with this structure no matter how R is allowed to vary.

So, in current orthodox population theory, to describe the entire range of population behaviors requires many different equations - each with its own specific justification. Exponential growth has a limited range of validity, as does the Logistics Equation, and any other equation of first order. Oscillatory behavior demands that a new paradigm be requisitioned; the Lotka-Volterra equations. And none of these describe population decline, nor extinction.

No structure exists that embraces - *in one single statement* - all possible behaviors. Contemporary theory offers no overriding principle that governs the gamut of population behaviors. To produce such a structure is the aim of what follows.

## 4. Conceptual foundations for an overriding structure

We seek a mathematical equation to embrace all of the great variety of population behaviors. The equation is built on some foundational axioms. Empirical verification of the equation they produce is what will measure the validity of these axioms. The axioms are:

*First: Variations in population number, n, are due entirely to environment.*

Conceptionally we partition the universe into two: the population under consideration and its environment. We assume that the environment drives population dynamics; that the environment is entirely responsible for time variations in population number – whether within a single lifetime or over many generations.

The survival and reproductive success of any individual is influenced by heredity *as well as* the environment it encounters. This statement doesn't contradict the axiom. The individual comes provisioned with heredity to face the environment. Both the environment and the population come to the present moment equipped with their capacities to influence each other; capacities derived from their past histories.

That the environment molds the population within a lifetime is clear; think of a tornado, a disease outbreak, or a meteor impact. That the environment governs population dynamics over generations is precisely the substance of 'natural selection' in Darwinian evolution.

That principle may be summarized as follows: "... the small selective advantage a trait confers on individuals that have it..." (Carroll 2006) increases the population of those individuals. But what does 'selective advantage' mean? It means that the favored population is 'selected' by the environment to thrive. Ultimately it is the environment that governs a population's history. Findings in epigenetics that the environment can produce changes transmitted across generations (Gilbert *et al.* 2009; Jablonka *et al.* 2009) adds further support to this notion.

Much productive research looks at traits in the phenotype that correlate with fitness or LRS (Lifetime Reproductive Success). (Clutton-Brock 1990; Coulson *et al.* 2006) The focus is on how the organism *fits* into its environment. So something called 'fitness' is attributed to the organism; the property of an organism that favors survival success. But environmental selection from among the available phenotypes is what determines evolutionary success. The environment is always changing so whatever genetic attributes were favorable earlier may become unfavorable later. Hence there is an alternative perspective: fitness, being a matter of selection by the environment, is induced by it and may, thus, be seen as a property of the environment.

Although, fitness, in some sense, is 'carried' by the genome, it is 'decided' by the environment. Assigning a fitness to an organism rests on the supposition of a static environment; one into which an organism fits or does not fit. A dynamic environment incessantly alters the 'fitness' of an organism.

This is the perspective underlying the axiom that *variations in population number, n, are due entirely to environment.*

In this view, although birth rates minus death rates yield population growth they are *not the cause* of population dynamics; rather birth and death rates register *the effect* of the environment on the population. This view parallels that of N. Owen-Smith who writes: ( Owen-Smith 2005) ".. population growth is not the result of a difference between births and deaths (despite the appearance of this statement in most text- books), but rather of the difference between rates of uptake and conversion of resources into biomass, and losses of biomass to metabolism and mortality"

*Second: An increasing growth rate is what measures a population's success.*

The 'success' of a population is an assertion about a population's time development; it concerns the size and growth of the population. A reasonable notion of success is that the population is *flourishing*. We want to give quantitative voice to the notion that flourishing growth reveals a population's success.

Neither population number, n, nor population growth, dn/dt, are adequate to represent 'flourishing'. Population number may be large but it may be falling. Such a population cannot be said to be flourishing. So we can't use population number as the measure of success. Growth seems a better candidate. But, again, suppose growth is large but falling. Only a rising growth rate would indicate 'flourishing'. This is exactly the quantity we propose to take as a measure of success; the growth in the growth rate. By flourishing is meant growing faster each year. That the rate of change of growth is a fundamental consideration in population dynamics has been advocated in the past. (Ginzburg and Colyvan 2005)

A corollary of these two foundational hypotheses is that *change is perpetual*. Equilibrium is a temporary condition. What we call equilibrium is a stretch of time during which $dn/dt = 0$. Hence 'returning to equilibrium' is not a feature of analysis in this model. "Biological persistence (is) more a matter of coping with variability than balancing around some equilibrium state." (Owen-Smith 2002)

Another corollary is this: The environment of one population is other populations. It's through this mechanism that interactions among populations occur: via reciprocity - if A is in the environment of B, then B is in the environment of A. So the structure offers a natural setting for 'feedback'. (Pelletier *et al.* 2009) It provides a framework for the analysis of competition, of co-evolution and of predator-prey relations among populations. These obey the same equation but differ only in the signs of coefficients relating any pair of populations.

**5. The Opposition Principle: quantitative formulation**

Based on the understandings outlined above we propose that an overriding principle governs the population dynamics of living things. It is this: *The effect on the environment of a population's success is to alter that environment in a way that opposes the success.* In order to refer to it, I call it the Opposition Principle. It is a *functional* principle (McNamara *et al.* 2009) operating irrespective of the *mechanisms* by which it is accomplished. In the way that increasing entropy governs processes irrespective of the way in which that is accomplished.

The Principle applies to a society of living organisms that share an environment. The key feature of that society is that it consists of a number, n, of members which have an inherent drive to survive and to produce offspring with genetic variation. Their number varies with time: n = n(t).

Because we don't know whether n, itself, or some monotonically increasing function of n is the relevant parameter, we define a *population strength*, N(n). Any population exhibits a certain strength in influencing its environment. This population strength, N(n), expresses the potency of the population in affecting the environment – its environmental impact.

N. Owen-Smith (Owen-Smith 2005) urges us to 'assess abundance' in terms of biomass rather than population. What is here named 'population strength' is related to that idea. Population number, itself, may not be a measure of environmental potency.

Perhaps this strength, N, is just the number n, itself. The greater n is, the more the environmental impact. But it takes a lot of fleas to have the same environmental impact as one elephant. So we would expect that the population strength is some function of n that depends upon the population under consideration. Biomass is one such function.

Two things about the population potency, N, are clear. First, N(n) must be a monotonically increasing function of n; dN/dn > 0. This is because when the population increases then its impact also increases. Albeit, perhaps not linearly. Second, when n=0 so, too, is N=0. If the population is zero then certainly its impact is zero. One candidate for N(n) might be n raised to some positive power, p. If p=1 then N and n are the same thing. Another candidate is the logarithm of (n+1).

We need not specify the precise relationship, N(n), in what follows. Via experiment it can be coaxed from nature. The only way that N depends upon time is parametrically through its dependence on n. In what follows we shall mean by N(t) the dependence N(n(t)). We may think of N as a surrogate for the number of members in the population.

The population strength growth rate, g = g(t), is defined by

$$g := \frac{dN}{dt} \qquad (3)$$

Like N, g too acquires its time dependence parametrically through n(t).
g = (dN/dn)(dn/dt)

To quantify how the environment affects the population we introduce the notion of 'environmental favorability'. We'll designate it by the symbol, f. It represents the effect of the environment on the population.

A population flourishes when the environment is favorable. Environmental favorability is what drives a population's success. We may be sure that food abundance is an element of environmental favorability so f increases monotonically with nutrient amount. It decreases with predator presence and f decreases with any malignancy in the environment - pollution, toxicity.

But in the last section we arrived at a quantitative measure of success. The rate of growth of the population strength - 'the growth of growth' or dg/dt – measures success. Hence, that a population's success is generated entirely by the environment can be expressed mathematically as:

$$\frac{dg}{dt} = f \qquad (4)$$

By omitting any proportionality constant we are declaring that f may be measured in units, (time)$^{-2}$. Since equation (4) says that success equals the favorability of the environment, it follows that f measures not only environmental favorability but also population success. One can gauge the favorability of the environment - the value of f - by measuring population success.

We're now prepared to caste the Opposition Principle as a mathematical statement. The Principle has two parts. 1. Any *increase* in population strength *decreases* favorability; the more the population's presence is felt the less favorable becomes the environment. 2. Any *increase* in the growth of that strength also *decreases* favorability.

Put formally: That part of the change in f due to an increase in N is negative. Likewise the change in f due to an increase in g is negative. Here is the direct mathematical rendering of these two statements:

$$\frac{\partial f}{\partial N} < 0 \quad \text{and} \quad \frac{\partial f}{\partial g} < 0 \qquad (5)$$

We can implement these statements by introducing two parameters. Both w and $\alpha$ are non-negative real numbers and they have the dimensions of reciprocal time. (Negative w values are permitted but are redundant.)

$$\frac{\partial f}{\partial N} = -w^2 \quad \text{and} \quad \frac{\partial f}{\partial g} = -\alpha \tag{6}$$

These partial differential equations can be integrated. The result is:

$$f = -w^2 N - \alpha g + F(t) \tag{7}$$

The 'constant' (with respect to N and g) of integration, F(t), has an evident interpretation. It is the gratuitous favorability provided by nature; the gift of nature. Equation (7) says that environmental favorability consists of two parts.

One part depends on the number and growth of the population being favored: the N and its time derivative, g. This part has two terms both of which always act to decrease favorability. These terms express the Opposition Principle.

The other part - F(t) - is the gift of nature. *There must be something in the environment that is favorable to population success but external to that population else the population would not exist in the first place.* This gift of nature may depend cyclically on time. For example, seasonal variations are cyclical changes in favorability. Or it may remain relatively constant like the presence of air to breathe. It may also exhibit random and sometimes violent fluctuations like a volcanic eruption or unexpected rains on a parched earth. So it has a stochastic component. All of these are independent of the population under consideration. In fact, however, dF/dt may depend on population number since this is the rate of consumption of a limited food supply.

Inserting equations (3) and (4) into (7) we arrive at the promised differential equation governing population dynamics under the Opposition Principle. It is this.

$$\frac{d^2N}{dt^2} + \alpha \frac{dN}{dt} + w^2 N = F(t) \tag{8}$$

In the world of physical phenomena this equation is ubiquitous. Depending upon the meaning assigned to N it describes electrical circuits, mechanical systems, the production of sound in musical instruments and a host of other phenomena. So it is very well studied. The exact analytical solution to (8), yielding N(t) for any given F(t), is known.

**6. Fits to empirical data.**

To explore some of the consequences of this differential equation we consider the easiest case; that the gift favorability is simply constant over an extended period of time. Assume F(t) = c independent of time. Non-periodic solutions arise if $\alpha \geq 2w$. As displayed in Figure 1 these produce results similar to the Verhulst equation - the Logistics equation. And like that equation they exhibit an exponential-like growth over a limited range.

Empirical data on such an exponential-like growth is exhibited in Figure 2 showing the population of musk ox (Ovibos moschatus) isolated on Nunivak Island in Alaska (Spencer and Lensink 1970). The data, gathered every year from 1947 to 1968, is in Table 1 of Spencer and Lensink's paper. It is, indeed, well fit by an exponential curve showing the formidable growth rate of 13.5% per year.

The population cannot possibly fit such a curve indefinitely. Nothing grows without end. The exponential curve fits the population data only in the domain shown. But, in this domain, the data is also well fit by an Opposition Principle curve where α= 0.02 per year, w = 0.0076 per year and n=N.

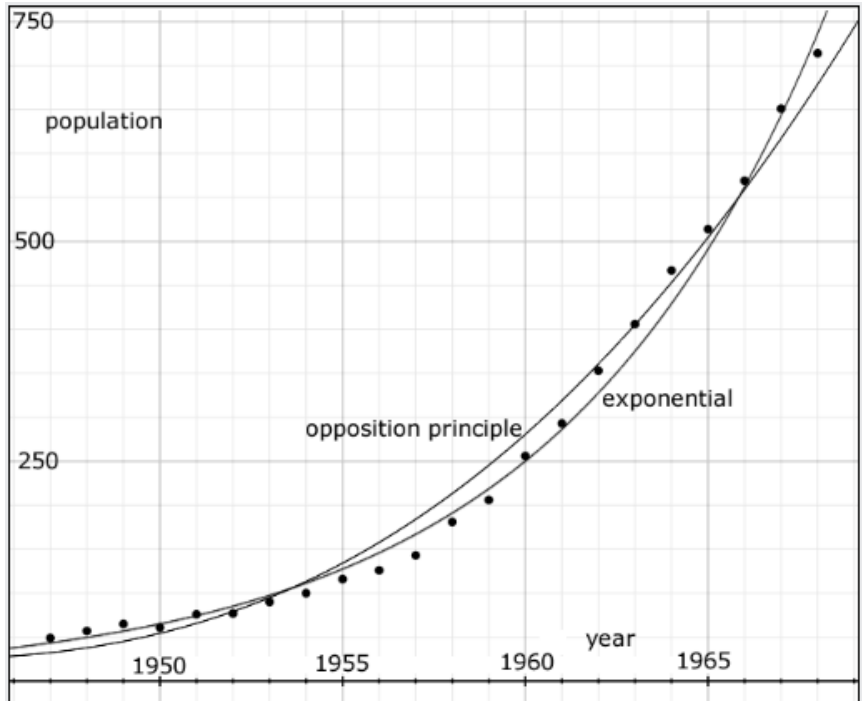

Figure 2. Data points, gathered every year from 1947 to 1968, reporting the number of musk ox on Nunivak Island, Alaska. The curves show that both an exponential and an Opposition Principle curve may be fit to the data.

If α < 2w the solutions to (8) are periodic and are given by:

$$N(t) = \frac{c}{w^2} + Ae^{-\alpha t/2} \sin(\omega t + a) \tag{9}$$

where the amplitude, A, and the phase, a, depend upon the conditions of the population at a designated time, say t=0. And the oscillation frequency, ω, is given by:

$$\omega := w\sqrt{\left(1-(\frac{\alpha}{2w})^2\right)} \qquad (10)$$

In Figure 3, equation (9) is compared to empirical data. The figure shows the population fluctuations of larch budmoth density (Turchin *et al.* 2003) assembled from records gathered over a period of 40 yrs. The data points and lines connecting them are shown in black. The smooth blue curve is a graph of equation (9) for particular values of the parameters.

We assumed α is negligibly small so it can be set equal to zero. The frequency, ω, is taken to be 2π/(9yrs) = 0.7per year. The vertical axis represents N. In the units chosen for N, the amplitude, A, is taken to be 0.6 and c is taken to be 0.6 per year². The phase, a, is chosen so as to insure a peak in the population in the year 1963; a=3.49 radians.

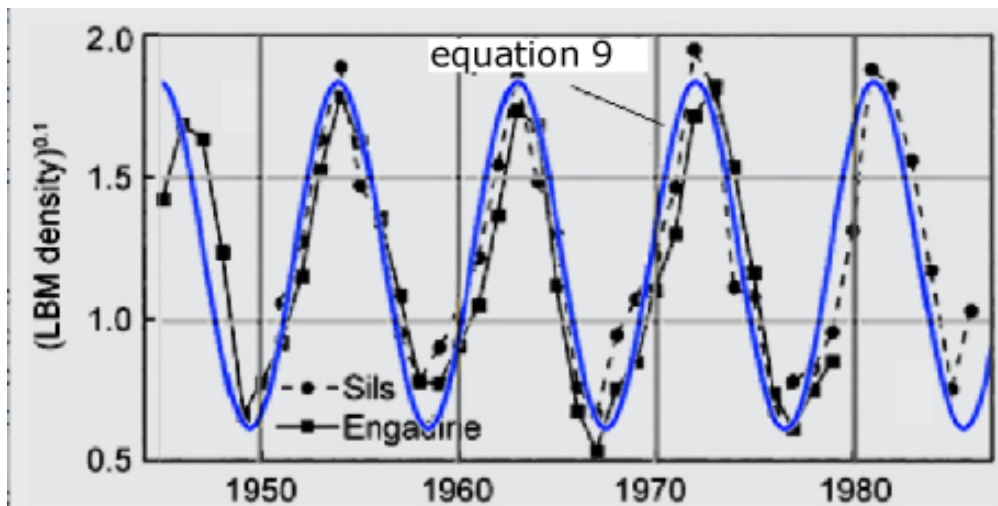

Figure 3. Observational data on the population fluctuations of larch budmoth density is shown as black circles and squares. The smooth blue curve is a solution of Equation (9).

Because the fluctuations are so large the authors plotted $n^{0.1}$ as the ordinate for their data presentation. The ordinate for the smooth blue theoretical curve is N. Looking at the fit in Figure 3, suggests how population potency may be deduced from empirical data. One is led to conclude that the population strength, N(n), for the budmoth varies as the 0.1 power of n. But the precision of fit may not warrant this conclusion.

The conclusions that may be warranted are these:

Considering that no information about the details of budmoth life have gone into the computation the graphical correspondence is noteworthy. It suggests that those details of budmoth life are nature's way of implementing an overriding principle. The graphical correspondence means that, under a constant external environmental favorability, a population should behave not unlike that of the budmoth.

Equation (9) admits of circumstances in which population extinction can occur. If $A > c/w^2$ then N can drop to zero. Societies with zero population are extinct ones. (On attaining zero, N remains zero. The governing differential equation, (8), doesn't apply when N<0.)
But the value of A derives from initial conditions; from N(t=0) and g(t=0). So depending upon the seed population and its initial growth rate the population may thrive or become extinct even in the presence of gift favorability, c. This result offers an explanation for the existence of the phenomenon of 'extinction debt' (Kuussaari *et al.* 2009) and a way to compute the relaxation time for delayed extinction.

The case explored reveals that periodic population oscillations can occur without a periodic driving force. Even a steady favorability can produce population oscillations.

Among the plethora of solutions to the governing differential equation, (8), is this one: Upon a step increase in environmental favorability – say, in nutrient abundance – the population may overshoot what the new environment can accommodate and then settle down after a few cycles.

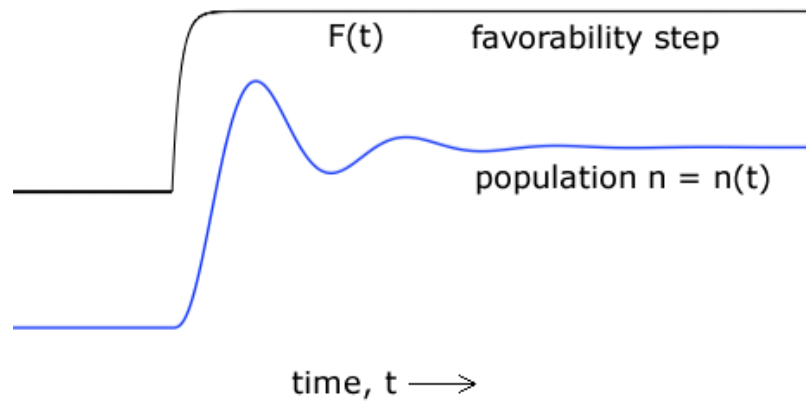

Figure 4. A theoretical population history that can result from the Opposition Principle differential equation.

Figure 4 illustrates this behavior. That there are such solutions amounts to a prediction that population histories like that of Figure 4 will be found in nature. In fact it has been found.

The data points (diamonds) of Figure 5 (Blount, Boreland, Lenski 2008) records the population of Escherichia coli (using optical density, OD, to measure it) maintained over 30,000 generations on a nutrient containing citrate which it could not exploit. Around generation 33,100 a mutation arose allowing a strain of the species to exploit this nutrient component. For those with this mutation the environment became suddenly more favorable and they flourished. Shown in the same figure is the theoretical curve arising from the Opposition Principle equation (8) with parameters $\alpha = 0.028$ per generation and $w = 0.023$ per generation and $n=N^{1.32}$.

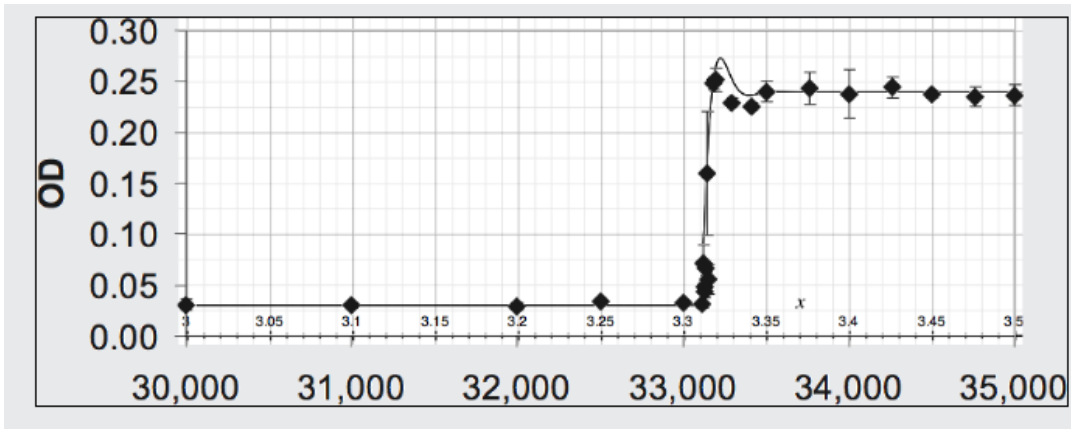

Figure 5. Fit of Opposition Principle curve to the data on a strain of E coli for which, because of a mutation, a step increase occurs in the favorability of its environment.

## 7. Conclusion

We noted at least five disparate regimes of population history - each with it's own individual and disjoint descriptive equation: exponential-like growth, saturated growth, population decline, population extinction, and oscillatory behavior. Another regime, without a theory, is population overshoot. It's argued here that these regimes can be brought under the embrace of a single differential equation describing them all.

That equation is the mathematical expression of general concepts about how nature governs population behavior. Being quantitative it offers us a framework with which to validate or refute these concepts. They are itemized as axioms and principles (Section 4). Some of these run counter to accepted convention thus making empirical refutation a substantive matter.

In short: a refutable proposition about the nature of populations is offered for assessment by the scientific community. Verification of the proposed equation would establish a basic understanding about the nature of living organisms.

McNamara, J.M. and Houston, A.I. 2009 Integrating function and mechanism. *Trends Ecol. Evol.* 24, 670-675

Murray, J.D., 1989 *Mathematical Biology*, Berlin: Springer-Verlag

Murray, B. G., Jr. 2000. Universal laws and predictive theory in ecology and evolution. Oikos 89, 403-408

Murray, B. G., Jr. 2001. Are ecological and evolutionary theories scientific? *Biological Reviews* 76, 255–289

Nowak, M. A. 2006 *Evolutionary Dynamics: Exploring the Equations of Life*. Canada: Harvard Press

O'Hara, R. B. 2005 The anarchist's guide to ecological theory. Or, we don't need no stinkin' laws. *Oikos* 110, 390-393

Okada, H., Harada, H., Tsukiboshi, T., Araki, M. 2005 Characteristics of Tylencholaimus parvus (Nematoda: Dorylaimida) as a fungivorus nematode. *Nematology* 7, 843-849

Owen-Smith, R. N. 2002, *Adaptive Herbivore Ecology*, Cambridge, Cambridge University Press

Owen-Smith, N. 2005 , Incorporating fundamental laws of biology and physics into population ecology: the metaphysiological approach. *Oikos* 111, 611-615

Parry, G.D. 1981 The meanings of r- and K-selection. *Oecologia* 48, 260-264
OppositionPrinciple    p. 26    January 1, 2012